Incorporation of Verifier Functionality in the Software for Operations and Network Attack Results Review and the Autonomous Penetration Testing System

Jordan Milbrath & Jeremy Straub
Institute for Cyber Security Education and Research
North Dakota State University
1320 Albrecht Blvd., Room 258
Fargo, ND 58108
Phone: +1-701-231-8196
Fax: +1-701-231-8255
Email: jordan.milbrath@ndsu.edu, jeremy.straub@ndsu.edu

**Abstract**

The software for operations and network attack results review (SONARR) and the autonomous penetration testing system (APTS) use facts and common properties in digital twin networks to represent real-world entities. However, in some cases fact values will change regularly, making it difficult for objects in SONARR and APTS to consistently and accurately represent their real-world counterparts. This paper proposes and evaluates the addition of verifiers, which check real-world conditions and update network facts, to SONARR. This inclusion allows SONARR to retrieve fact values from its executing environment and update its network, providing a consistent method of ensuring that the operations and, therefore, the results align with the real-world systems being assessed. Verifiers allow arbitrary scripts and dynamic arguments to be added to normal SONARR operations. This provides a layer of flexibility and consistency that results in more reliable output from the software.

**1. Introduction**

The software for operations and network attack results review (SONARR) is designed to perform security assessment on mission critical systems, which cannot be taken offline or risk damage from conventional penetration testing [1]. The autonomous penetration testing system (APTS) builds on the SONARR architecture to actually conduct automatic penetration testing [2].

Verifiers are used within the SONARR system to update networks' fact values to ensure that they correspond with the real-world entities that they represent. As is typical with the concept of digital twins [3], the purpose of SONARR network facts is to mirror the real-world entities that they represent. Being able to update facts in real-time means that the network will be able to provide relevant data and interact with live systems in a way that is both accurate and effective. Verifiers are associated with facts to verify single facts, or they can be associated with common properties to provide a more general verification capability.

Verifiers work by running a script that retrieves the real-world value that corresponds with a given fact. For example, if there is a fact that represents whether or not a computer is connected to the internet, then a script would be run that checks this and then returns true or false, accordingly. The result of this script is returned to the software, the facts is updated, and network processing operations continue as normal. Verifiers prevent the network from significantly diverging from the real-world entities it was designed to represent and draw conclusions about.

Without verifiers, SONARR has no way to determine if its representation of facts is accurate, as compared to the real-world entities that they are designed to represent. SONARR does not have the ability to retrieve any information from its environment, which may be changing, or may start in a state that was not anticipated. Verifiers cause SONARR to adapt to its environment and allow it to provide correct results that account for unforeseen changes and dynamic environments.

This paper introduces and evaluates the addition of verifiers to SONARR and APTS. This paper continues with Section 2, which discusses prior work on which the current work builds.  Then, Section 3 describes the changes that were made to SONARR to support verifiers in their current state and then talks about how they function conceptually. Following this, Section 4 provides more details on how verifiers work in the current implementation of SONARR, using an example to explain. Section 5 describes and analyzes testing that was completed to evaluate the efficacy of verifiers. It also provides detail about the practical benefits of including verifiers in SONARR. Finally, the paper concludes with Section 6 which describes the steps needed to further integrate verifiers into the SONARR infrastructure and some areas that could make the use of verifiers even more helpful for users.

## 2. Background

This section discusses areas of prior work related to the addition of verifiers to SONARR. First, the Blackboard Architecture is reviewed. Then, work related to discovery in automated penetration testing is discussed.

### 2.1 Blackboard Architecture

In the 1960s and 1970s, expert systems were introduced with the creation of Dendral [4] and Mycin [5], two systems that used rule-fact structures for decision making. The Blackboard Architecture was first established in 1985 by Hayes-Roth [6]. Notably it contributed new ideas to the control flow of previous expert systems. The Blackboard Architecture, more recently, has added the concept of actions, which allow the system to directly influence its environment [7]. In addition to this, key concepts related to generic-ification and reuse of rules and separating logical and physical / organizational relationships (including common properties, containers, and links) have been added to the functionality of the Blackboard Architecture [8], [9].

The Blackboard Architecture's adaptability allows it to be used in many application areas, and its expert-system roots makes it a highly explainable form of artificial intelligence (AI).  The technology has been used for work with unmanned system failure detection [10], medical image interpretation [11], terrorism countering [12], and legal decision making [13], where detailed reporting of the AI's inner workings is required. Other areas that have made use of the Blackboard Architecture include software testing [14], cybersecurity attack modeling [15], recognizing handwriting [16], sound identification [17], creating mathematical proofs [18], and even generating poetry [19].

### 2.2 Discovery in Penetration Testing

Penetration testing involves approaching a system from the perspective of an attacker in an attempt to find vulnerabilities in that system that could be exploited [20]. The National Institute of Standards and Technology (NIST) has defined a four-stage penetration testing methodology [21]. The steps involved in this methodology are planning, discovery, attack, and reporting. During the discovery step, penetration testers collect information about the system they are testing. This step is crucial because it provides information that can be used in the attack and reporting steps. One key part of this discovery phase is scanning. With passive scanning, data is collected without system interaction. Active scanning, on the other hand, requires the system to be acted upon and information to be collected from it and returned to the penetration tester [22]. The information that is retrieved from a scan will then be used during the attack phase to decide on attack techniques that will be used and to supply arguments for attacks that are run [21].

In the context of automated penetration testing tools, the discovery phase uses automated scanning tools to collect information and return it to the tool, similar to how the penetration tester would typically collect data [23]. Having this data is critical for the operations of the automated penetration testing tool for the same reasons that it is critical for manual penetration testers: it allows them to configure their attacks to the environment that they operate in. Penetration testers (and the automated system) can use this information to specifically create attacks and respond to the environment as it is influenced [24].

The current implementation of APTS does not have any discovery capabilities, so it skips this step during its operations. Verifiers work to fill this gap and allow the system adapt to its environment, thus filling the role that is required for the tool to be comprehensive and meet the standards set by NIST and other organizations that have worked with automated penetration testing tools in the past.

### 3. Additions Made to SONARR to Support Verifiers

In order for verifiers to be added to SONARR, several modifications were required. First, a verifier class needed to be added. This class includes several properties that specify how verifiers are executed. In order for a verifier process to execute, a file name must be provided and arguments may be provided. A verifier stores the executable path, which is provided as the process' file name. This value should correspond with the script that it runs. For example, if the verifier works through a PowerShell script, then the executable path should be 'PowerShell'. If it uses a Python script, it should be 'Python'. This value is plugged directly into the ProcessStartInfo object, so it is necessary for the system to be able to find the executable through the specified value. Thus, one should be able to open up a terminal, type in the executable path, and have the process execute.

The next property addition is the format string field. The format string specifies the format in which the arguments should be presented in the verifier command. In many cases, it is necessary for a specific script to be run. This script name will be specified as one of the arguments. How it is presented to the main executable as an argument may different depending on that executable. Some formats may need "-File" or "-Script" before the script name, while others may not need to specify an argument name at all and simply place the path of the script in that location. Having the format string as a property allows the format to be configurable for each individual verifier and specified specifically for each executable that will need to parse it.

An array of format arguments is used with each format string. These arguments specify the values that will be retrieved from a fact that is being verified. These arguments are based on the specific SONARR implementation and can be set to any arbitrary value. The verifier will look for a custom property whose common property has a description identical to the specified format argument. The description of that custom property will be the string that is used in the arguments for the executable. A concrete example is now described to clarify and describe this logic in more detail.

For example, a container object could be used to represent a computer. That container has a fact that indicates whether or not the computer is connected to the internet. In order to verify that it is connected to the internet, the computer pings an IP address. If the ping is successful, it concludes that it is connected to the internet: however, if the ping is unsuccessful, it concludes that it is not connected to the internet. A potential argument for this situation is the IP address that the computer pings, or its "ping target". Figure 1 shows the relationships and necessary information for a verifier to generate arguments for this ping request.

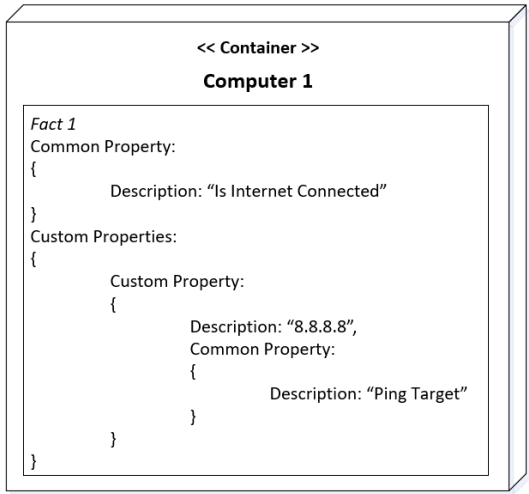

*Figure 1. Ping example for verifier arguments.*

If a verifier has a format argument of "Ping Target" and is given the "Fact 1", as shown in Figure 1, then the verifier will look at the fact's custom properties. In this case, there is only one. It finds the custom property whose common property has a description of "Ping Target" and retrieves the description of that custom property, which in this example is "8.8.8.8". When the verifier generates its argument string, it will use the value of "8.8.8.8" where the "Ping Target" should be placed.

There is one special case where the format argument is not retrieved from a fact in this way. This is when the format argument is "Description". When that is the case, instead of looking at the custom properties, the description of the fact is returned. This is relevant in scenarios where the description of a fact is necessary for the script that is retrieving its value.

A small test network of three contains was generated to demonstrate verifiers. It is shown in Figure 2.

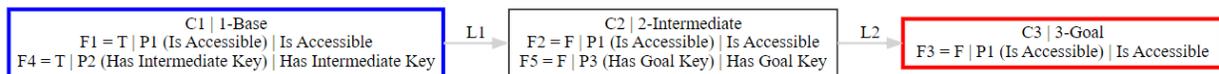

*Figure 2. Small test network of three containers.*

A single verifier script was created; however, depending on the environment and what that script returns, the results of the run may be different. As shown in the diagram above, container 1's "Has Intermediate Key" fact is set to true. A rule exists indicating that if this fact is true, traversal may proceed to container 2. The same logic applies to container 2 and its goal key. Initially, however, the "Has Goal Key" fact value is set to false. This relationship is similar to needing a key to move on to the next room in a hallway. If one has a key, they may proceed; otherwise, they must stop.

In this network, Fact 5 has a verifier that checks whether the correct goal key exists. In the test implementation, the verifier simply checks a file for contents that match a target value.

## 4. Current Implementation

The current implementation of verifiers was designed as a proof of concept to display their functionality. It only supports a fixed set of verifiers, but it does so in a way that allows the easy interchanging of these verifiers for testing purposes. The executable path and format string are hardcoded and cannot be readily changed by the user, but they can easily be changed by a testing developer. The fact, custom property, and common property values can be readily changed by users through the SONARR interface, and these changes may affect network traversal.

An example container with verifier arguments is shown in Figure 3. The example verifier's executable path is "PowerShell", and its format string is:

*-ExecutionPolicy Bypass -File "Verifier1.ps1" "{0}" "{1}"*

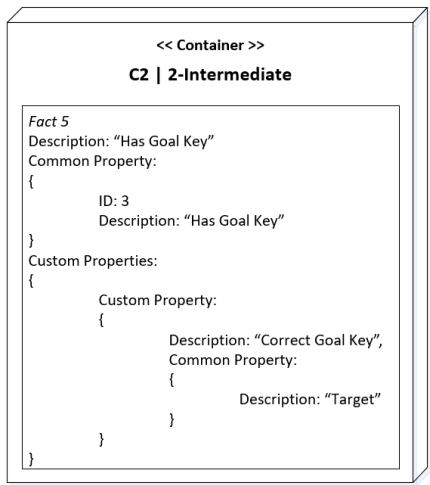

*Figure 3. Container 2's verifier arguments.*

The format arguments, for this verifier, are "Description" and "Target". Using the format string above, the value retrieved for "Description" will go in place of {0}, and "Target" will go in place of {1}. Fact 5's description is used as the value for "Description" as described previously. To determine the value for "Target", the custom properties for Fact 5 are examined. There is a custom property that has a common property with the description "Target", so the description of that custom property is used for the second argument. In this case, that value is "Correct Goal Key". Given this, the command that will be run, as a new process is:

*PowerShell -ExecutionPolicy Bypass -File "Verifier1.ps1" "Has Goal Key" "Correct Goal Key"*

Whatever is returned from this process run is parsed as a bool and the value of Fact 5 is set to this. The example verifier script is shown in Listing 1.

*Listing 1. Verifier script.*

```powershell
param([Parameter(Mandatory=$true)]$folderName,
[Parameter(Mandatory=$true)]$targetKeyValue)

$targetFolderFullName = Join-Path -Path $PSScriptRoot -ChildPath $folderName
$childItems = Get-ChildItem -Path $targetFolderFullName
$targetFileContent = Get-Content -Path $childItems[0].FullName

return $targetFileContent -ceq $targetKeyValue
```

For testing purposes, this verifier script checks the content of the first file within the specified folder. If the contents of that file match the target value, it returns true; otherwise, it returns false.

The file that the verifier script checks is on the system that runs the script, and it is not influenced by SONARR traversal processing. It is only read from. To test the efficacy of the implementation, traversals were run.

The first run was done with the key file's contents set to "Correct Goal Key", which allows the traversal to continue to the goal node. During this run, the verifier PowerShell script was run and returned "True", allowing a path to be generated from container 1 to container 3.

For the second run, the key file's contents were set to "Not Correct Goal Key". This traversal did not result in any paths being found, and debug information indicates that the PowerShell script returned a value of "False", thus stopping traversal.

In addition to demonstrating the correct functionality of the verifier software implementation, this shows that using verifiers, traversal may verify fact values that are dependent on environmental or external variables. This will affect the results of traversal, according to the external state and values retrieved.

**5. Testing, Analysis, and Impacts**

This section describes the testing process that was used to analyze the performance of verifiers and their impact on the overall function of SONARR. The testing process and speed impact is described and analyzed in Section 5.1. Then, the implications of varying verifier timing are assessed in Section 5.2. Finally, the practicality of verifiers and some potential use cases are described in Section 5.3.

**5.1. Performance Impact**

Testing was conducted to determine the impact of verifiers on the performance of the SONARR and APTS systems. The system was tested under three different scenarios. First, the system was run with no verifiers. Then, it was run using the PowerShell script verifier described in Section 4. Finally, he system was run using a Windows batch script verifier. There were 1,000 testing iteration completed for each scenario, with 3,000 iterations completed total. A single testing iteration involved a full APTS traversal of the network as described in Section 4. Each testing iteration was timed, and the average time elapsed for each experimental condition is shown in Figure 5. This testing demonstrates the impact of a very basic verifier to demonstrate the computational performance it would have.  Most actual verifiers would have a more pronounced impact on speed, due to their comparatively greater functionality and, thus, greater processing and other requirements.

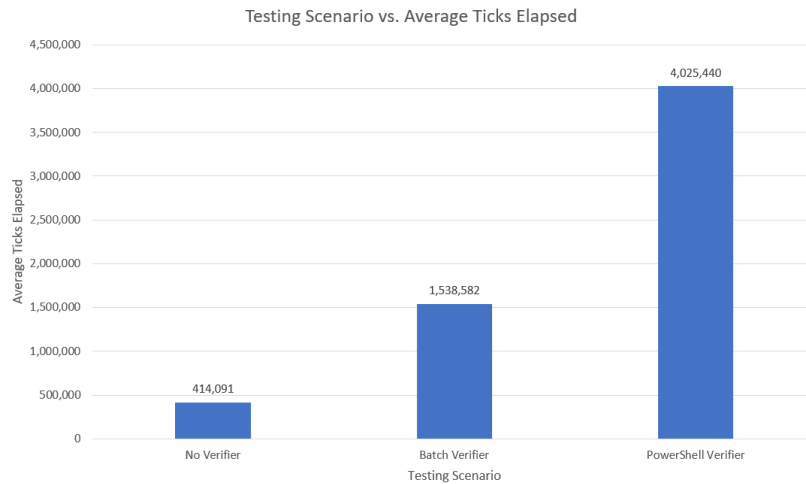

*Figure 5. Testing scenario vs. average ticks elapsed.*

The testing indicated that including verifiers in APTS traversal processing increases the time that it takes to complete that traversal. In the case of a batch file, which is a relatively low-level type of script, it took an average of 1,538,582 ticks to complete, which is 1,124,490 more than the average when no verifier was used. This means that including a batch verifier increased the overall traversal time by approximately 271%. While batch verifiers are useful in many scenarios for simple operations, PowerShell or similarly more complex scripts may be more practical to use to verify fact values. During the tests, it was found that, on average, runs that included PowerShell verifiers took 872% longer to complete than those without any verifiers.

Part of the reason for the large increase in time when running verifiers is due to the need to create a new process to run the verifier. Creating a new process takes a notable amount of time, relative to the rest of the operations that are completed during traversal. As shown by the difference between the batch verifier and the PowerShell verifier's times, the method of verification also impacts the time elapsed. PowerShell scripts offer more capabilities under many scenarios, and they provide easier ways to complete these complex tasks; however, they take notably longer than batch files to run.

This speed impact is important to consider when creating verifiers for networks. Depending on the desired speed of traversal, processing the verifier scripts and the executables that those scripts use to operate may need to be optimized. Optimization of verifier scripts may help to ensure that APTS traversals run smoothly and in as little time as possible. Keeping verifiers as light-weight as possible will help to ensure that they complete their processing in as little time as possible.

**5.2. Accuracy**

The APTS system, using the verifiers, verified the desired fact values accurately 100% of the time, during testing. This was to be expected, due to the nature of the verifiers used.

The impact that using verifiers has on traversal should be taken into consideration while developing a network and scenario for that network to operate in. While not using verifiers significantly reduces the time spent, even for the lightweight scripts such as batch scripts, in scenarios where the environment is not completely controlled, fact values may not always be in the state that the system expects them to be in. Including verifiers in traversals allows operators to ensure that the system is producing correct results (that parallel the state of the target system). Verifiers also help the system to be more adaptable in scenarios where the environment may change frequently. Ensuring that the APTS network is updated consistently allows the results of each run to be accurate at the time of traversal.

Since the operating environment may change, APTS may generate different results when run with the same arguments at different points during runtime. Since verifiers change the values of facts in the APTS network, this may cause different rules to be triggered, resulting in different actions firing. This may cascade, in some scenarios, into very different end states being reached within that environment. An end state reached without verifiers is, thus, not indicative of the state that will be reached with verifiers. Because of changing information, it is possible that a state completely different is reached, when verifiers are used.

Runtimes of verifiers may differ, quite significantly in some scenarios. In networks that have facts that change values over time, the speed of the verifier that is used may impact the fact values that are retrieved. For example, consider a scenario where there are two verifiers, verifier A and verifier B, and verifier A takes one second to retrieve a fact's value, while verifier B takes two seconds to retrieve that same fact's value. This could result in different outcomes. If the fact's value changes after 1 second and before 2 seconds, between when Verifier A would capture the fact value and Verifier B would, then the two verifiers, which are both designed to retrieve the same fact value, would return different results for the same fact due to the speed at which they get that fact value. Considerations like this, related to verifiers, must be analyzed when constructing networks, creating verifiers, and analyzing results of APTS runs. Ensuring that all aspects of verifiers and the implications of using them are understood is key to using them effectively.

## 5.3. Practicality Impacts

The practical benefits of verifiers are described in this section. Section 5.3.1 details how verifiers reduce the likelihood of results that inaccurately represent their real-world counterparts. Section 5.3.2 describes how verifier scripts can greatly extend the functionality of SONARR and APTS and how their dynamic argument support allows complex scripts to be used during traversal.

### 5.3.1. Fixing Inaccurate Results

Prior to verifiers, APTS had the ability to run actions and influence the environment in which it is running; however, it lacked the ability to update or receive input back from its environment. This limited its ability to provide up to date output for its user. In many cases, users of APTS will not have all of the information necessary to fully predict the outcome of actions in the execution environment. Because of this, when actions are run, they may result in different results then were predicted. Verifiers fill this need and are thus crucial to the robustness of APTS. This is demonstrated by the results presented in Section 4.

To illustrate the impact of adding verifiers to APTS, the execution environment was set up to make a real-world traversal from the entity represented by container 2 to the entity represented by the end container (container 3), shown in Figure 2, impossible. This change was not, however, mirrored in the fact configuration of the imported network. This represents a scenario when the user does not know everything about the execution environment or the environment changes after a network is created.

Figure 6 shows the result of running the network without the inclusion of verifiers. It returns one path, which asserts that APTS was able to make it from the start container to the end container. This path was returned because there were not any verifiers to look at the execution environment and recognize that that the fact values that were previously loaded do not correctly represent the execution environment. Thus, this path is inaccurate, since the execution environment made this traversal impossible. This means that APTS returned a path that could not truly be completed in the real world. Without verifiers, it is implicitly assumed that all of the information APTS was provided is accurate and that the environment did not change, which was not the case for this particular run.

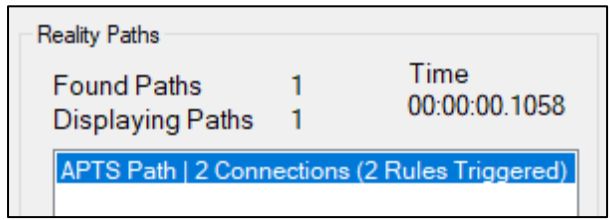

*Figure 6. Result of impossible APTS run without verifiers.*

There are several reasons that providing inaccurate results could be problematic. First, showing results that are inaccurate will be confusing for the end user. For example, the end container may not have been reached in the executing environment, and the fact values displayed in the APTS interface may not match the real world. This may result in networks in APTS that are substantially different from what they are intended to represent. Figure 7 shows the correct representation of the network after an APTS run with verifiers for "Has Intermediate Key" (Fact 4) and "Has Goal Key" (Fact 5), where the end container was not reached. Figure 8 shows the network run without verifiers. It is shown that Facts 2, 3, and 5 all have different values in the network run without verifiers than the network that represents the environment correctly. As networks get larger, the impact of inaccurate fact values has the potential to grow tremendously.

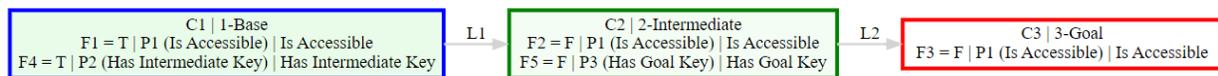

*Figure 7. Correct representation of the testing network (with verifiers)*

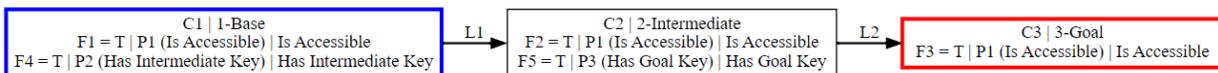

*Figure 8. Incorrect representation of the testing network (without verifiers).*

In addition to providing an inaccurate representation to the end user, making assumptions during traversal could create unintended consequences in the executing environment. While this test used passive actions that did not notably impact the environment, scenarios that use more powerful tools could result in unintended damage to the system and place it in a state that inhibits or alters its functionality, deletes necessary files, or creates other unintended issues. For larger networks, the level of divergence and negative impact may be proportionally larger.

It is vital that the APTS system knows when the environment is not responding to actions as expected, and that it responds to that. Without this ability, APTS would simply generate a fixed set of actions that it would run, based solely off of its input data without any consideration of the current state of the environment that it resides in. Verifiers connect APTS to its environment, allowing APTS to generate results that correctly represent that environment at present. Figure 9 shows the result of an APTS run in the impossible testing environment with the inclusion of verifiers.

Figure 9 shows that no complete paths are generated, which mirrors the real-world environment. In an environment such as this, where verifiers stop a traversal that would otherwise execute without their inclusion, actions that may alter or harm the system (due, for example, to the APTS representation of the network not matching the real-world) are not executed. This prevents issues from arising due to inaccuracy. It also provides the end user with results that correctly represent the state of the environment.

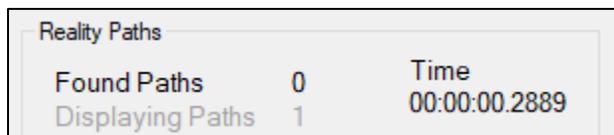

*Figure 9. Result of impossible APTS run with verifiers*

*5.3.2. Dynamic Argument Support and Complex Scripting*

Since verifiers pull arguments for their executables from custom properties, facts may have many arguments that are specific to them. Pairing this idea with customized verifiers allows a wide range of capabilities.

This flexible structure allows for complex verifiers to be developed. For example, if a fact is created to represent whether the executing environment has access to a computer, that the fact could have a common property that references the computer. How that computer is referenced can be changed depending on what information is available and a verifier script can be created to match. If a computer name is known, a script can be created to ping computers by name. If an IP address is known, a script can be created to ping computers by IP address. If there is a file containing the computer's name on a server that the executing computer has access to, a script may be written that retrieves that file, reads and parses the text, and then pings the computer based on that. A single verifier script could also be created that takes any of these arguments as an input, checks which type of argument was provided, and pings the computer using the appropriate technique.

To enable this, any parameters within verifier scripts that are fact-specific should be placed into arguments that are stored within custom properties specific to each fact that uses the verifier. This allows several facts to use the same verifier. The number of verifiers necessary for a given set of facts is dependent on the flexibility of the verifier script itself, as scripts can be made to account for different formats of arguments.

When SONARR networks are created to represent real-world computer networks, verifiers can be used to run scripts that remotely assess or which are automatically loaded on computers that are not the original executing computer. For example, a script could be created that takes a target computer reference (such as an IP address) as an input. A second input could be the filename of a script that will be sent to and run on the target computer. The first script may gain access to the computer with the given IP address and send over the contents of the second script to run. The details of this execution would be dependent on the script itself. Languages, such as Python or PowerShell, are capable of tasks such as this. Verifiers provide a dynamic foundation that supports complex scripting, allowing for complex procedures that can be completed in a script to be triggered, with arguments.

## 6. Conclusions and Future Work

This paper has described and evaluated a functional implementation of verifiers. It has demonstrated the capabilities that verifiers provide and characterized their impact on SONARR and APTS system performance. Additional work is needed to allow users to easily work with verifiers; however, the work described herein provides the foundation to build upon.

The limitations of the current implementation of verifiers are described in this section along with a discussion regarding the future work needed. Section 6.1 describes the addition of the verifier files and file support to match other network object types that currently exist within SONARR. Then, Section 6.2 describes the addition of verifier information to the user interface, allowing users to see the results of verifier scripts and how they are affecting traversal.

### 6.1. Extending Verifier Type Support

Several additions will be needed for the current infrastructure of SONARR and APTS to allow verifiers to be used seamlessly. First, verifiers will need to be imported with networks. This means that they will need their own files that include an ID, executable path, format string, and format arguments. This is necessary for the verifiers to determine which facts they should pull data from to execute their commands.

The current implementation allows users to supply arbitrary custom property values and therefore arbitrary arguments for the verifier processes that were used for testing; however, as a proof of concept, it does not allow custom verifiers to be imported themselves. Thus, the system could be further expanded by having files that store information about verifiers, allowing custom ones to be user-defined, and allowing those files to be imported and associated with facts and common properties. This would facilitate a greater integration of verifiers into networks. Users could create their own verification scripts using any language that can be run from a terminal. Format strings and arguments also could be configured, for any scenario, allowing for prospectively any type of fact values to be collected.

Additionally, common properties will need a "verifiers" column added to their file. This will allow the verifier with a given ID to be associated with a common property. This connection allows verifiers to be executed during traversal. These changes will require alteration of the SONARR importing functions; however, the methods used to implement this new type of network object will be similar to those currently used.

### 6.2. Verifier Result Display

In many cases, especially where actions do not produce their intended outcomes, it may be beneficial for the end user to know which actions executed successfully and which did not. Verifiers validate their results against the intended values for each fact, so this result data could be made available to the user through the SONARR user interface. This would provide the user with a better understanding of what occurred during a given APTS run. This may allow the user to make changes to the network representation or alter the actions to produce the desired results.

**Acknowledgements**

Thanks is given to Cameron Kolodjski for his development work on the SONARR and APTS systems prior to this project. Thanks is also given to Matthew Tassava for his work on the SONARR software and to Anthony DeFoe for his work on visualization development for the SONARR software. Finally, thanks is given to other members of the project team for their feedback, testing and other contributions.